\documentclass[11pt]{article}
\usepackage[english]{babel}
\usepackage{amsmath,amsfonts,amssymb,amscd}

\usepackage{epsf}

\title{The Geometrical Modelling of Fluids}
\author{Mikhail G. Ivanov\thanks{e-mail:~\tt mgi@mi.ras.ru}\\
\it Moscow institute of physics and technology,\\
\it 141700, Institutskii per. 9, Dolgoprudny, Moscow Region,
Russia}
\date{May 3, 2009}
\begin{document}

\maketitle

\begin{abstract}
The paper considers the nonlinear electrodynamics type model and
its relation with relativistic hydrodynamics with no dissipation
(including string and membrane hydrodynamics). We are able to
convert arbitrary flux of fluid to the family of geodesics by the
conformal transformation of metric. The conditions of
transformation of nonlinear electrodynamics solution to linear
electrodynamics solution by changing of metric are presented.
\end{abstract}
\vspace{5mm}

\section{Nonlinear electrodynamics-type model}

  Let us consider the following action for the model
  of non-linear electrodynamics type with $(D-n-1)$-form
  potential $I$
  ($I_{M_1\dots M_{D-n-1}}=I_{[M_1\dots M_{D-n-1}]}$)
  in $D$-dimensional space-time with metric $g_{MN}$
  of signature $(-,+,+,\dots,+)$
\begin{equation}
  S_1[I]=\int d^Dx\sqrt{|g|}\,L(\|J\|),
  \label{action-ed}
\end{equation}
  where $J=dI$.

  We use the following notation:
\begin{eqnarray*}
   (dA)_{M_0M_1\dots M_k}=(k+1)\,\partial_{[M_0}A_{M_1\dots
   M_k]},\\
   (A,B)=\frac1{k!}A_{M_1\dots M_k}B^{M_1\dots M_k},\quad
   \|A\|^2=(A,A),\\
   (\delta A)^{M_1\dots M_{k-1}}=\frac1{\sqrt{|g|}}
   \partial_{M_k}\left(\sqrt{|g|}\,A^{M_1\dots M_{k-1}M_k}\right).
\end{eqnarray*}

  The \textquotedblright first pair of Maxwell
  equations\textquotedblright{} $dJ=0$
  is the consequence of the definition $J=dI$.

  The variation of action with respect to $I$ provides
  the field equations
  (\textquotedblright second pair of Maxwell equations\textquotedblright)
\begin{equation}
  \delta\left(J\,\frac{L'(\|J\|)}{\|J\|}\right)=0,\quad\Leftrightarrow
  \quad
  \partial_{M_{D-n}}\left(\sqrt{|g|}\,
  J^{M_1\dots M_{D-n}}\,\frac{L'(\|J\|)}{\|J\|}\right)=0.
\label{eom1}
\end{equation}

  The models becomes linear if $L(\|J\|)\sim\|J\|^2$.
  In this case field equations are $\delta J=0$.
  It is the regular Maxwell electrodynamics in 4-dimensional
  space-time if $D=4$, $n=2$.

\section{Relativistic hydrodynamics}

  The non-linear electrodynamics-type model (\ref{action-ed})
  could also describe the potential nondissipative flux
  of fluid, consists of $(n-1)$-dimensional
  membranes.
  The fluid is continuous distribution of membranes, where
  each point of $D$-dimensional space-time belongs to
  the single membrane world-surface $\mathbf{V}$.
  If $n=1$ the membrane is particle, i.e. we have a regular
  fluid.
  If $n=2$ the membrane is a string, i.e. we have a string-fluid.

  $(D-n)$-form $J$ is dual density of membrane flux.
  For regular fluid of particles ($n=1$)
  $J_{M_2\dots M_D}=j^{M_1}\,\sqrt{|g|}\,\varepsilon_{M_1\dots M_D}$,
  where $j^M$ is particle flux density, and
  $\varepsilon_{M_1\dots M_D}$
  ($\varepsilon_{M_1\dots M_D}=\varepsilon_{[M_1\dots M_D]}$,
  $\varepsilon_{01\dots D-1}=+1$)
  is totally antisymmetric symbol.

  $J$ specifies $(D-n)$ directions orthogonal to world-surface
  of dimension $n$, i.e. every vector $v^M$, tangent to
  world-surface $\mathbf{V}$, satisfies the condition
  $v^{M_1}J_{M_1\dots M_{D-n}}=0$.
  $\|J\|$ is the density of partiacles/strins/membranes,
  and \mbox{$-L(\|J\|)$} is energy density in attendant frame.

  {\bf Statement 1.} To describe the flux of
  particle/strings/membranes by field $J$ one specifies
  the world-surfaces as a level-surfaces for set of scalars
  $\varphi^\alpha$, $\alpha=1\dots,D-n$ (see \cite{0412318} and references therein).
  The field $J$ represented in terms of $\varphi^\alpha$ by the following equation
\begin{eqnarray}
\label{J-J}
  &&J=J_\varphi=f(\varphi)\,d\varphi^1\wedge\dots\wedge d\varphi^{D-n}
  ~\Leftrightarrow~\\
  &&~\Leftrightarrow~J_{M_1\dots M_{D-n}}=f(\varphi)\,(D-n)!\,\partial_{[M_1}\varphi^1\dots
  \partial_{M_{D-n}]}\varphi^{D-n},
  \nonumber
\end{eqnarray}
  where $f(\varphi)$ is a function of $\varphi^\alpha$.
  $f(\varphi)$ could be set to unit by redefinition of fields
  $\varphi^\alpha$, so we assume \mbox{$f(\varphi)\equiv1$}.

  The equation (\ref{eom1}) becomes the equation
  of potential flux of membrane fluid
  (regular fluid if $n=1$),
  consist of non-intersecting membranes with
  world-surfaces $\varphi=const$.
  Pressure acting on membrane in attendant frame is
$$
  p_\bot=L-\|J\|\cdot L'.
$$
  E.g. the Lagrangian $L(\|J\|)\sim\|J\|$ describes zero pressure
  membrane fluid (\textquotedblright membrane dust\textquotedblright).

  Similar to regular fluid, membrane fluid equations of motion
  could be derived from energy-momentum tensor continuity conditions
  $\nabla_M T^{MN}=0$.
  So, to prove the correspondence one has to check
  the energy-momentum tensor
$$
  T_{MN}=L(\|J\|)\,P_{MN}+p_\bot\,(g_{MN}-P_{MN}),
$$
 where
$$
P_{MN}=g_{MN}-\frac{J_{MM_2\dots M_{D-n}}J_{N}{}^{M_2\dots M_{D-n}}}{(D-n-1)!\,\|J\|^2}.
$$
  If $J$ is defined by equation (\ref{J-J}), then the tensor $P_{MN}$
  is a projector to tangent subspace at world-surface $\mathbf{V}$
  in the point considered.

   {\bf Statement 2.} One can skip the potentiality
   condition for the flux (\ref{J-J}) by the modification
   of action by $(D-n)$-form Lagrange factor $K$:
$$
  S_2[I,\varphi,K]=\int
  d^Dx\sqrt{|g|}\left[L(\|J\|)+(K,J-J_\varphi)\right].
$$
  After the removing of Lagrange factor $K$ new equations of motion
  could be represented in the following form
 (forms $J$ and $J_\varphi$ are identified):
\begin{equation}
 J_{M_1\dots M_{D-n-1}K}\,\,  \partial_{M_{D-n}}\left(\sqrt{|g|}\,
  J^{M_1\dots M_{D-n-1}M_{D-n}}\,\frac{L'(\|J\|)}{\|J\|}\right)=0.
\label{eom2}
\end{equation}
  Equations (\ref{eom2}) could be derived from equations (\ref{eom1})
  by projection to directions orthogonal to membrane world-surface.
  So, for the field (\ref{J-J}) equation (\ref{eom2})
  is more weak, than equation (\ref{eom1}).
  It corresponds to skipping of flux potentiality condition.

\section{Metric transformations and nonlinearity}
\subsection{Conformal transformations of metric}
  One can remove the extra factor (in comparison to linear theory)
  $\frac{L'(\|J\|)}{\|J\|}$ in field equations
  (\ref{eom1}), by modification of space-time metric.
  Let us consider the conformal transformation of metric:
\begin{eqnarray*}
  g_{MN}&\to&\tilde g_{MN}=F^2g_{MN},\\
  g^{MN}&\to&\tilde g^{MN}=F^{-2}g^{MN},\\
  \sqrt{|g|}&\to&\sqrt{|\tilde g|}=F^D\sqrt{|g|},\\
  J_{M_1\dots M_{D-n}}&\to&J_{M_1\dots M_{D-n}},\\
  J^{M_1\dots M_{D-n}}&\to&\tilde J^{M_1\dots M_{D-n}}
  =F^{-2(D-n)}\,J^{M_1\dots M_{D-n}},\\
  \|J\|&\to&\|\tilde J\|=F^{-(D-n)}\|J\|.
\end{eqnarray*}

  {\bf Statement 3.}
  If the conformal factor $F$, satisfies the condition
\begin{equation}
  F^{2n-D}=\frac{L'(\|J\|)}{\|J\|},
\label{transform-ed}
\end{equation}
  then using metric $\tilde g_{MN}$ field equations
  could be represented in the following form
$$
  \tilde\delta\tilde J=0.
$$
   I.e. in new metric one has standard linear equations
  for free (with no sources) closed $(D-n)$-form $J$.
   For the regular non-linear electrodynamics in
  4-dimensional space-time ($D=4$, $n=2$)
  the transformation (\ref{transform-ed}) is
  not possible.

   In the general case (if $D\not=2n$) one can use the conformal
 transformations of metric in both directions, connecting
 the solutions of different non-linearity in
 conformally-equivalent spaces.

    For linear electrodynamics-type model in space with any fixed
 metric superposition principle holds. According the principle
 sum of two field equation solutions is a new solution
 of the same equations.
    One could spread the superposition principle to nonlinear
 models using the following procedure.

  {\bf Statement 4.} {\bf Nonlinear superposition principle.}
   The field is described by action (\ref{action-ed}) with $D\not=2n$.
   Let $J^{(1)}$ and $J^{(2)}$ are solutions of field equations
  at some conformally-equivalent spaces with metrics
  $g^{(1)}_{MN}$ and $g^{(2)}_{MN}$, such that the metrics become
  the same after the transfer to linear model, i.e.
  $$
    F_{(1)}^2\,g^{(1)}_{MN}=F_{(2)}^2\,g^{(2)}_{MN}=g^{(0)}_{MN},
  $$
  where
  $$
    F_{(i)}=\left(\frac{L'(\|J^{(i)}\|_{(i)})}{\|J^{(i)}\|_{(i)}}\right)^{\frac1{2n-D}},
  $$
  here $\|\cdot\|_{(i)}$ is calculated using the metric $g^{(i)}_{MN}$.
   One can transform both solutions to solutions in the space with
  the same metric $g^{(0)}_{MN}$ and build a linear combination of
  transformed solutions:
\begin{equation}
    J^{(\alpha1+\beta2)}=\alpha\,J^{(1)}+\beta\,J^{(2)},
    \label{superpos-J}
\end{equation}
   $\alpha$ and $\beta$ are constants.
   The new field $J^{(\alpha1+\beta2)}$ is a solution
  of the same nonlinear equations at the space with new metric
\begin{equation}
   g^{(\alpha1+\beta2)}_{MN}=F^{-2}_{(\alpha1+\beta2)}g^{(0)}_{MN},
   \label{superpos-g}
\end{equation}
  where $F_{(\alpha1+\beta2)}$ is specified by the relation
 $$
  F_{(\alpha1+\beta2)}
  =\left(\frac{L'(\|J^{(\alpha1+\beta2)}\|_{(\alpha1+\beta2)})}{\|J^{(\alpha1+\beta2)}\|_{(\alpha1+\beta2)}}\right)^{\frac1{2n-D}}
  =\left(\frac{L'(F^{-(D-n)}_{(\alpha1+\beta2)}\,\|J^{(\alpha1+\beta2)}\|_{(0)})}{F^{-(D-n)}_{(\alpha1+\beta2)}\,\|J^{(\alpha1+\beta2)}\|_{(0)}}\right)^{\frac1{2n-D}}.
 $$

\subsection{Metric transform for
 \textquotedblright electrostatics\textquotedblright{} and
 \textquotedblright magnetostatics\textquotedblright}

   In special case $D=2n$ one could not remove nonlinearity by
  conformal metric transform.
   Nevertheless, if there is no time dependence, then the problem
  could be treated in the important cases of static fields in static
  space-time.

   Projection of $(D-n)$-form $J$ to surface $t=const$ produce
  $(D-n)$-form $J^{H}$ (components are numerated by indices with no time)
  and $(D-n-1)$-form $J^E$ (components are numerated by indices with
  time):
$$
  J=J^H+dt\wedge J^E.
$$
  If $D=2n=4$, then form $J^E$ is covector of electric field,
  and $J^H$ is 2-form of magnetic field.

  If $J^H=0$ (\textquotedblright electrostatic\textquotedblright{} case)
  or $J^E=0$ (\textquotedblright magnetostatic\textquotedblright{} case),
  Lagrangian is a function of лагранжиан
  $\|dt\wedge J^E\|$ or $\|J^H\|$.

  Let $g_{0\alpha}=g^{0\alpha}=0$, for $\alpha\not=0$.
  In this case $\|dt\wedge J^E\|=\sqrt{g^{00}}\,\|J^E\|_h$,
  where $\|\cdot\|_h$ is calculated using the metric
  $h_{\alpha\beta}=g_{\alpha\beta}$ ($\alpha,\beta=1,\dots,D-1$),
  restricted to $(D-1)$-dimensiona space $t=const$.

  Now one can introduce two $(D-1)$-dimensional actions.
\begin{equation}
  S_H[I_H]=\int d^{D-1}x\sqrt{|h|}\,L_H(\|J^H\|_h),\quad
  L_H(\|J^H\|_h)=\sqrt{g_{00}}\,L(\|J^H\|_h),
  \label{action-H}
\end{equation}
\begin{equation}
  S_E[I_E]=\int d^{D-1}x\sqrt{|h|}\,L_E(\|J^E\|_h),\quad
  L_E(\|J^E\|_h)=\sqrt{g_{00}}\,L(\sqrt{g^{00}}\,\|J^E\|_h),
  \label{action-E}
\end{equation}
  where $J^E=dI_E$, $J^H=dI_H$.
  Here $I_H$ is $(D-n-1)$-form
  (for regular electromagnetic field it is vector-potential),
  and $I_E$ is $(D-n-2)$-form
  (for regular electromagnetic field it is scalar potential).

  So, instead of one theory we study two static theories
  at space $t=const$.
  Each of the theories admits the conformal transform described above
  (\ref{transform-ed}), which makes theory linear.
  Now the transform is not space-time transform, but pure space transform.
  After the transformations one has linear
  \textquotedblright electrostatics\textquotedblright{}
  and linear \textquotedblright magnetostatics\textquotedblright.

 {\bf Statement 5.} For \textquotedblright electrostatics\textquotedblright{}
 and \textquotedblright magnetostatics\textquotedblright{}
 nonlinear superposition principle similar to (\ref{superpos-J}), (\ref{superpos-g})
 is applicable with replacement of parameters of action (\ref{action-ed})
 to appropriate parameters of action (\ref{action-E})
 or action (\ref{action-H}).

\subsection{Other metric transformations}
   In this section we consider the useful examples of metric transformations,
  which are not conformal.

  {\bf Example 1.}
  Metric of non-rotating black hole could be replaced
  by flat metric of Minkowski space-time, if linear
  electrostatic field is considered.
  The metric has the form
\begin{equation}
  ds^2=k(r)\,dt^2-\frac{dr^2}{k(r)}-r^2\,(d\theta^2+\sin^2\theta\,d\varphi^2).
  \label{bh-g}
\end{equation}
  Electrostatic potential has the only component $I_r$, i.e. $I=a(r)\,dt$.
  Electric field also has the only component
  $J_{rt}$, i.e. $J=a'(r)\,dr\wedge dt$.
  Omitting of $k(r)$ in $g_{tt}$ and $g_{rr}$ does not change
  $g=\det(g_{MN})$ and the only contravariant field component $J^{rt}$.
   So, one can replace in the electromagnetic field equations
  black hole metric (\ref{bh-g}) by Minkowski space-time metric.
  In the example considered we transform linear electrodynamics
  to linear electrodynamics again in different space-time.
   The transformation considered is based upon the specific form
  of metric tensor. The conformal transformations we discussed
  above were applicable to arbitrary metric of certain dimension.

  {\bf Example 2.} Similarly, spherically-symmetric solution
  of nonlinear electrodynamics equations $J=a'(r)\,dr\wedge dt$
  in spherically-symmetric space-time
$$
  ds^2=k(r)\,dt^2-q(r)\,dr^2-r^2\,(d\theta^2+\sin^2\theta\,d\varphi^2)
$$
  could not be transformed to linear theory by conformal
  transformation ($D=2n=4$).
  Nevertheless, we can transform radial and time components
  of metric to transform the particle-like solution into
  the vacuum solution of linear theory:
$$
  g_{tt}=k(r)\to F^2 g_{tt},\qquad
  g_{rr}=-q(r)\to F^{-c} g_{rr}.
$$

\section{Metric transformations in hydrodynamics}

   Let field $J$ specifies the solution of (membrane) fluid.
   The model with zero pressure $p_\bot=0$ is (membrane) dust.
   In dust model particles (membranes) do not interact with each other.
   This model could be considered to be preferable.
   For (membrane) dust lines/surfaces of flow $\varphi=const$ in space-time
 are geodesics lines/surfaces.

  {\bf Statement 6.} If we choose the factor $F$
  according to the following condition
  (instead of condition (\ref{transform-ed}))
\begin{equation}
  F^n=L'(\|J\|),
\label{transform-dust1}
\end{equation}  then in the metric $\tilde g_{MN}$ the equation of motion
  has the form
\begin{equation}
  \tilde\delta\left(\frac{\tilde J}{\|\tilde J\|}\right)=0.
\label{eom-dust1}
\end{equation}
  The equation describes (membrane) dust with the dual density
 of particles/strings/mem\-branes flow $\tilde J$.

   If one applies the transformation (\ref{transform-dust1})
  to electrostatics or magnetostatics, then
  the world surfaces of electric field-lines or of
  magnetic field-surfaces are geodesic surface of a metric
  conformally-equivalent to initial one.
   For electrostatics in Minkowski space-time
$$
  F^4\sim \mathbf{E}^2.
$$
  It transforms the point charge into infinite tube.

   {\bf Example 3.} For single particle in Minkowski space-time
$$
  ds^2=dt^2-dr^2-r^2(d\theta^2+\sin^2\theta\,d\varphi^2)
  \quad\to\quad
  d\tilde s^2=\frac{dt^2}{r^2}-\left(\frac{dr}r\right)^2-(d\theta^2+\sin^2\theta\,d\varphi^2).
$$

  Electrostatic fields does not depend upon time.
  So, it is natural to represent field-lines as geodesics
  of 3-dimensional geometry (not as sections $t=const$ of
  geodesics surfaces of 4-dimensional geometry).

  If one removes time coordinate, it produces
  the theory with $D=3$, $n=1$ (electrostatics)
  and the theory with $D=3$, $n=2$ (magnetostatics).

  So, we get the other (3-dimensional) conformal factor,
  which makes electric field-lines geodesics:
$$
  F^2\sim \mathbf{E}^2.
$$
  This transformation revert charges \textquotedblright upside
  down\textquotedblright{}
  (see the next example).

   {\bf Example 4.} For the single charge Euclidean space undergo
   the inversion transformation:
$$
  dl^2=dr^2+r^2(d\theta^2+\sin^2\theta\,d\varphi^2)
  \quad\to\quad
  d\tilde l^2=\left(d\frac1r\right)^2+\left(\frac1r\right)^2(d\theta^2+\sin^2\theta\,d\varphi^2).
$$

  In the magnetostatic case $D=3$, $n=1$ the transformation
  is generated by factor
$$
  F^4\sim \mathbf{H}^2.
$$
   Here, since the field-surfaces belongs to
  subspace $t=const$, the conformal factor is the same
  in 3-dimensional and 4-dimensional cases.

\section{Conclusion}

  Many authors, working in general relativity,
  postulate, that the theory combines two styles:
  \textquotedblright high\textquotedblright{} geometrical
  style of Einstein tensor and \textquotedblright low\textquotedblright{}
  field style of energy-momentum tensor.
  The geometrical interpretation of matter field allows
  to make field style \textquotedblright higher\textquotedblright{},
  i.e. more geometrical.

  The main result of the paper is the demonstration of possibility
  to modify matter action by metric transformations.
  In many cases one can transform nonlinear theory
  into liner one and introduce analogue of superposition
  principle.
  Nevertheless nonlinearity does not disappear,
  it is hidden in the metric transformation.

  Similarly to many other problems, the most physical
  case (electromagnetic field in 4-dimensional space-time)
  is exceptional case, linear and nonlinear electrodynamics
  could not be transformed to each other by conformal
  metric transformation due to conformal invariance.

  The reduction of relativistic dynamics of (membrane) fluid
  to description of family of geodesics is the other interesting
  application of conformal metric transformation.
  So, the pressure of membranes/strings/particles is modified
  (it could be set to zero) by conformal transformations.
  The pressure provides the interaction between the close
  membranes. If the pressure is switched off, then
  one has just free motion of nonintersecting
  membranes/strings/particles.
  It could be interesting for string and brane theories
  (see review \cite{strings} and references therein)
  and for relativistic elasticity theory
  \cite{0412318,elastic}.

  Description of dynamics through the introduction of the additional
  metric is also considered in
  {\it acoustic geometry}, i.e. it describes
  {\it acoustic black holes}
  \cite{unruh}
  (large number of publications could be found
  by the key words).
  In this approach the wave equation in non-relativistic fluid
  is written in terms of some pseudo-Riemannian metric.
  For fluid in rest it is the Minkowski metric with
  speed of light replaced by speed of sound.

\section*{Acknowledgements}
    The author is grateful to M.O. Katanaev for useful discussion.
    The author is grateful to all organizers and participants
  of International Conference on Mathematical Physics and Its Applications
  (Samara, September 8–-13, 2008) for possibility to present
  and discuss the results in warm and friendly atmosphere.

    The work was partially supported by grants
   RFFI-08-01-00727 and NSh-3224.2008.1.

\end{document}